\documentclass[prl,aps,reprint]{revtex4-1}

\usepackage{amsthm} 
\usepackage{amsmath}
\usepackage{hyperref}
\usepackage{latexsym}
\usepackage{amsfonts}
\usepackage{amssymb}
\usepackage{mathtools} 
\usepackage{color}
\usepackage{bbm,dsfont}
\usepackage{graphicx}
\usepackage{ifthen}
\usepackage{enumerate}
\usepackage{textcomp}

\usepackage[T1]{fontenc}
\usepackage[utf8]{inputenc}

\newtheorem{theorem}{Theorem}

\theoremstyle{definition}

\newtheorem{remark}{Remark}

\newcommand{\R}{\mathbb{R}} 

\newcommand{\abs}[1]{\left| #1 \right|} 
\newcommand{\avg}[1]{\left\langle #1 \right\rangle} 
\newcommand{\ket}[1]{\left| #1 \right\rangle} 
\newcommand{\braket}[3][ ]{\left\langle #2 \vphantom{#3} \middle| #3 \vphantom{#2} \right\rangle_{#1}} 
\newcommand{\matrixel}[3]{\left\langle #1 \vphantom{#2#3} \right| #2 \left| #3 \vphantom{#1#2} \right\rangle} 
\newcommand{\tr}[2][]{\mathrm{tr}_{#1}\left[#2\right]} 
\let\baraccent=\= 
\renewcommand{\=}[1]{\stackrel{#1}{=}} 

\newcommand{\cD}{\mathcal{D}}

\newcommand{\sfe}{{\sf E}}
\newcommand{\sff}{{\sf F}}

\newcommand{\Eo}{\mathsf{E}}
\newcommand{\Fo}{\mathsf{F}}
\newcommand{\Go}{\mathsf{G}}
\newcommand{\Ho}{\mathsf{H}}
\newcommand{\Mo}{\mathsf{M}}
\newcommand{\No}{\mathsf{N}}
\newcommand{\Qo}{\mathsf{Q}}

\newcommand{\var}[1]{\mathrm{Var}(#1)}
\newcommand{\cov}[1]{\mathrm{Cov}(#1)}

\newcommand{\ketbra}[2]{\left| #1 \vphantom{#2} \middle\rangle \middle\langle #2 \vphantom{#1} \right|}

\begin{document}

\title{Focusing in Arthurs-Kelly-type Joint Measurements with Correlated Probes}

\author{Thomas J Bullock}
\email{tjb525@york.ac.uk}
\affiliation{\small Department of Mathematics, University of York, York, UK}

\author{Paul Busch}
\email{paul.busch@york.ac.uk}
\affiliation{\small Department of Mathematics, University of York, York, UK}

\date{\small  \today}

\begin{abstract}
Joint approximate measurement schemes of position and momentum provide us with a means of inferring pieces of complementary information if we allow for the irreducible noise required by quantum theory. One such scheme is given by the Arthurs-Kelly  model, where information about a system is extracted via indirect probe measurements, assuming separable uncorrelated probes. Here, following Di Lorenzo (PRL {\bf 110}, 120403 (2013)), we extend this model to both entangled and classically correlated probes, achieving full generality. We  show that correlated probes can produce more precise joint measurement outcomes than the same probes can achieve if applied alone to realize a position or momentum measurement. This phenomenon of {\em focusing} may be useful where one tries to optimize measurements with limited physical resources. Contrary to Di Lorenzo's claim, we find that there are no violations of Heisenberg's error-disturbance relation in these generalized Arthurs-Kelly models. This is simply due to the fact that, as we show, the measured observable of the system under consideration is covariant under phase space translations and as such is known to obey a tight joint measurement error relation.
\end{abstract}

\maketitle

\emph{Introduction.}\quad The incompatibility of the position and momentum observables is a well-known feature of quantum mechanics and is succinctly expressed by the preparation uncertainty relation \cite{Kennard27,Weyl}:
\begin{equation}
	\var{Q,\psi}\var{P,\psi}\geq\frac{\hbar^2}{4},\label{eq:PUR}
\end{equation}
which states that for any state $\psi$ that we prepare a system in, the product of the variances in the statistics of the position $Q$ and momentum $P$ is bounded below by Planck's constant. 
However, while \eqref{eq:PUR} is well understood as a preparation uncertainty relation, there is controversy over how this incompatibility may be expressed when we consider measurements of \emph{both} observables on the same system. Heisenberg \cite{Heisenberg27}
formulated a trade-off relation for the error $\Delta Q$ of a position measurement and the resulting disturbance $\Delta P$ of momentum:  
\begin{equation}
\Delta Q\,\Delta P\geq \frac{\hbar}{2},\label{eq:"Heisenberg"}
\end{equation}
which he obtained on the basis of heuristic arguments. In recent years this form of tight bound has been called into question by some
\cite{Ozawa2004,Branciard2013,DiLorenzo2013} and corroborated by others \cite{BLW2013c,DN2014,WSU2011}.
This controversy is the result of a lack of universally agreed upon operational definitions of error and disturbance. 

Here we analyze the work of Di Lorenzo \cite{DiLorenzo2013}, whose claim of a violation of \eqref{eq:"Heisenberg"} results from the consideration of a particular measurement model and a specific choice of measures of error and disturbance. His scheme, a generalization of the celebrated model of Arthurs and Kelly \cite{AK65}, couples a system to two probes that are then measured to provide approximate information about both the position and momentum of the system. The suggested extension consists of allowing for initial correlations between the two probes. We  show that the purported violation of \eqref{eq:"Heisenberg"} does not occur in the most general extension of the Arthurs-Kelly model.

In what follows, we consider the generalized Arthurs-Kelly model and allow for correlation between the probes. We derive the effective joint observable measured on the system, which is represented operationally as a positive operator valued measure (POVM) on phase space. It is found that, for arbitrary preparations of the probes, this observable is covariant under phase space translations; i.e., such translations do not change the observable but instead shift its associated probability distributions. 
This generalizes the case of uncorrelated probes, where the covariance property has previously been shown \cite{Busch1985}. 

Covariant phase space observables are known to  always satisfy an error-error relation of the form \eqref{eq:"Heisenberg"} \cite{BLW2013c}. Given that a sequential measurement is a form of joint measurement, with the disturbance corresponding to the approximation error in the second observable (see, e.g., \cite{Ozawa2004,BLW2013c,BHL}), it follows that error-disturbance relations are special cases of error-error relations. 
That the error and disturbance measures given in \cite{DiLorenzo2013} lead to a violation of \eqref{eq:"Heisenberg"}  suggests something is wrong with these measures. 

In fact, we show that Di Lorenzo's ``disturbance'' is actually a measure of the relative imprecision of two approximations of the ideal position or momentum observable. The first approximation is the marginal observable derived from the Arthurs-Kelly model, while the second is the one measured by a single probe. As Di Lorenzo found, this relative imprecision can indeed  become negative. This observation leads to an interesting phenomenon that we refer to as  {\em focusing}: the marginal observables in a joint measurement can be more precise than the observables measured by the individual probes. Focusing may prove useful for improving the performance of measurements with limited physical resources.

We give two cases where both the approximate position and momentum observables are focused by performing a joint measurement. In one instance the probes are prepared in a pure entangled state, while in the second they are prepared in a separable mixed state. This latter case shows that entanglement does not help over classical correlations. 

In what follows, we will consider three particles with one continuous degree of freedom. Each particle is described by either a pure state belonging to the Hilbert space $L^2(\R)$ of square-integrable complex functions over the real line $\R$ or by a density operator. Further to this, we will set $\hbar=1$ for simplicity.

\begin{figure}[floatfix]
	\includegraphics[width=.45\textwidth]{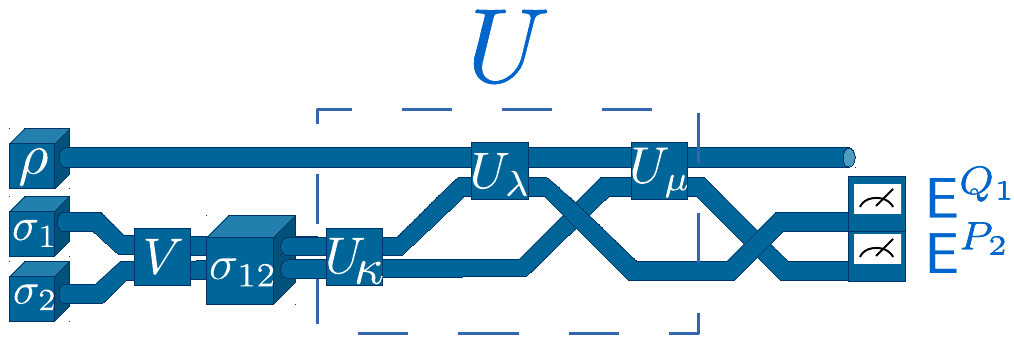}
	\caption{\label{fig:full} An extension of the Arthurs-Kelly model to allow for correlated states. Two probes, which are coupled together by a unitary $V$, are then coupled to the system described by state $\rho$ by a unitary $U$. After this coupling, ideal measurements of the position of the first probe and momentum of the second are performed, from which we infer information about the position and momentum of our considered system.}
\end{figure}
\emph{The model.}\quad We consider an extension of the Arthurs-Kelly model \cite{AK65}, as shown in Fig. \ref{fig:full}. The model couples a system, described by a state $\rho$, to two probes, labeled 1 and 2, via an impulsive (short-timed) unitary interaction $U=\exp(-iH)$, where $H$ is the interaction Hamiltonian
\begin{equation}
	H= \lambda Q P_1 -\mu P Q_2 + \tfrac{\lambda\mu}{2}\kappa P_1Q_2.
\end{equation} 
The numbered operators $P_1$ and $Q_2$ refer to the momentum and position operators on probes 1 and 2, respectively, whil the unnumbered operators $Q$ and $P$ are the position and momentum operators, respectively, on the considered system. The positive coupling constants $\lambda$ and $\mu$ determine the strength of the coupling between the two probes and the system, while $\kappa$ determines the coupling strength between the two probes. Using Baker-Campbell-Hausdorff decompositions  of $U$:
\begin{align}
\begin{split}
	U&=\exp(i\mu P Q_2)\exp(-i\lambda Q P_1) \exp[-i\tfrac{\lambda\mu}{2}(\kappa-1) P_1Q_2]\\
	&= \exp(-i\lambda Q P_1)\exp(i\mu P Q_2)\exp[-i\tfrac{\lambda\mu}{2}(\kappa+1) P_1Q_2],\label{eq:unitary}
\end{split}
\end{align}
we see that the joint measurement can alternatively be implemented as a sequence of interactions and measurements. If $\abs{\kappa}=1$, this can be considered as a strictly sequential measurement of position and momentum (with the ordering depending on the sign of $\kappa$). These coupling constants are assumed large enough that we may ignore the free evolution of the system and the probes.
After this coupling, the two probes are ideally measured, i.e., we perform projection-valued measures on the probes. The first probe has its position measured by $\Eo^{Q_1}$, and the second probe its momentum by $\Eo^{P_2}$. From the statistics of these measurements, we infer information about the position and momentum of the considered system. In most works, the probes were assumed to be in pure, uncorrelated states; i.e., they are described by a product state $\varphi_1\otimes\varphi_2$, say, with $\varphi_i$ the state of the $i$th probe. Assuming such a situation for the probes, the model produces an effective joint observable $\Go$ on the considered system that is covariant under phase space translations \cite{Busch1985}. By covariance we mean that the application of a phase space translation $W_{qp}=\exp[-i(qP-pQ)]$ to the POVM element $\Go(Z)$, with $Z\subseteq \R^2$ an interval in phase space, will result in another element of the same POVM, but with a shifted input value $Z+(q,p)$:
\begin{equation}
	W_{qp}\Go(Z)W^*_{qp}=\Go(Z+(q,p)).
\end{equation}
Instead of the pure product state $\varphi_1\otimes\varphi_2$ describing the probes, we consider a generally correlated state $\sigma_{12}$, which may be seen as the result of a unitary coupling $V$; i.e., $\sigma_{12}=V(\sigma_1\otimes\sigma_2)V^*$, as in Fig. \ref{fig:full} \cite{Note2}. In the case $\sigma_{12}$ is a pure entangled state, $\sigma_{12}=P_{\varphi_{12}}$ (the projector onto the normalized vector $\varphi_{12}$), we find coupling the probes to our system initially in a pure state $\rho=P_\psi$ produces the state of the combined system in the position representation:
\begin{align}
	\Psi(q,q_1,q_2) &=	U(\psi\otimes\varphi_{12})(q,q_1,q_2)\\
	&=\psi(q+\mu q_2)\varphi_{12}(q_1-\lambda q - \tfrac{\lambda\mu}{2}(\kappa+1)q_2,q_2),\notag
\end{align}
where $q$, $q_1$ and $q_2$ are the position coordinates for the system and probes 1 and 2, respectively.
Similarly, if $\sigma_{12}$ is mixed, we consider the mixed state $U(P_\psi\otimes\sigma_{12})U^*$. From this, the probability of finding the pointer readings belonging to the intervals $X,Y \subseteq \R$, respectively, is given by $\braket{\Psi}{I\otimes \Eo^{Q_1}(X)\otimes \Eo^{P_2}(Y)\Psi}$. With appropriate scaling of the intervals, this probability can be interpreted as the likelihood of finding the system, described by state $\psi$, belonging to the phase space cell $X\times Y \subseteq \R^2$:
\begin{equation}\begin{split}
	\braket{\Psi}{I\otimes \Eo^{Q_1}(\lambda X)\otimes \Eo^{P_2}(\mu Y)\Psi}\label{eq:prob-rep}\equiv \braket{\psi}{\Go^{(\lambda,\mu)}(X\times Y)\psi}.
\end{split}
\end{equation}
(Here $\lambda X=\{\lambda x|x\in X\}$, and similarly for $\mu Y$.) The positive operators $\Go^{(\lambda,\mu)}(X\times Y)$ are mathematically defined by \eqref{eq:prob-rep} and form a phase space observable on the system.
We can now state our 
main result
\cite{Note3}.
\begin{theorem}
The observable $\Go^{(\lambda,\mu)}$ given by an Arthurs-Kelly model with arbitrary probe state is a covariant phase space observable.
\end{theorem}
This extends what has been considered in the past to all possible probe preparations for the Arthurs-Kelly model and shows that the results found for probes prepared in pure product states can be readily generalized.

Covariant phase space observables have been studied thoroughly elsewhere (see, e.g., \cite{Holevo1979,Werner1984,Cassinelli2003,KLY2006a}), and it is well known that any such observable $\Go$ may be expressed in the form
\begin{equation}
	\Go(Z) = \Go_\tau(Z) = \frac{1}{2\pi}\int_Z dq\,dp\, W_{qp} \tau W_{qp}^*,
	\label{eq:cov-obs-gen-form}
\end{equation}
where $Z\subseteq \R^2$ and $\tau$ is a unique positive operator with unit trace; i.e., mathematically, $\tau$ is a density operator. The marginal observables of $\Go$ are approximations of the ideal position and momentum operators. In particular, the marginals of $\Go^{(\lambda,\mu)}$, denoted by $\Eo^{(\lambda,\mu)}$ and $\Fo^{(\lambda,\mu)}$, are 
\begin{subequations}
\begin{align}
	\label{eq:marginala}\Eo^{(\lambda,\mu)} (X) = \Go^{(\lambda,\mu)}(X\times \R) = (\chi_X * e^{(\lambda,\mu)})(Q),\\
	\label{eq:marginalb}\Fo^{(\lambda,\mu)} (Y) = \Go^{(\lambda,\mu)}(\R\times Y) = (\chi_Y * f^{(\lambda,\mu)})(P),
\end{align}
\end{subequations}
where $\chi_A$ denotes the indicator function onto the subset $A\subseteq \R$, $*$ denotes convolution, and both $e^{(\lambda,\mu)}$ and $f^{(\lambda,\mu)}$ are probability distributions. These probability distributions characterize the noise in the measurement statistics of $\Eo^{(\lambda,\mu)}$ and $\Fo^{(\lambda,\mu)}$ and depend on the state $\sigma_{12}$ of the probes and the coupling constants. Owing to the structure of $\Go$ as given in \eqref{eq:cov-obs-gen-form}, these distributions are identical to the probability distributions of position and momentum in the state represented by $\tau_-$,
the space inversion of $\tau$. Hence it is evident that their variances obey the standard uncertainty relation
\begin{equation}\label{eq:ur}
\var{e^{(\lambda,\mu)}}\,\var{f^{(\lambda,\mu)}}\ge\frac14.
\end{equation}
The second moment of (say) the distribution $e^{(\lambda,\mu)}$ is expressed in terms of the first moment (the mean) and the variance via
\begin{equation}
	e^{(\lambda,\mu)}[2]=e^{(\lambda,\mu)}[1]^2+\var{e^{(\lambda,\mu)}}.\label{eq:e2}
\end{equation} 
This can be interpreted as representing both systematic and random error contributions inherent in $\Eo^{(\lambda,\mu)}$ as an approximation of the ideal position measurement $\Eo^Q$. This intuition is strengthened by the fact that one can find physically relevant measures of the error, $\Delta$, of approximating an ideal observable by another, such that $\Delta(\Eo^{(\lambda,\mu)},\Eo^Q)^2=e^{(\lambda,\mu)}[2]$ \cite{Note3}. 
The inequality 
\begin{equation}
	e^{(\lambda,\mu)}[2]f^{(\lambda,\mu)}[2]\ge 1/4,
\end{equation} 
obtained as a direct consequence of \eqref{eq:ur},
is therefore an instance of the joint measurement uncertainty relation \eqref{eq:"Heisenberg"}. This result holds for all covariant phase space observables, including the ones that arise from our extension of the Arthurs-Kelly model. 

As we shall show,  Di Lorenzo's claim of a violation of \eqref{eq:"Heisenberg"} is a result of an inadequate definition of disturbance.

\emph{Di Lorenzo's disturbance.}\quad We first consider the model of the previous section and set one of the coupling constants, either $\lambda$ or $\mu$, to zero. If we put $\mu=0$, the coupling unitary $U$ reduces to $U_\lambda=\exp(-i\lambda Q P_1)$, and we measure with just the first probe. Similarly, if we set $\lambda=0$, then $U$ reduces to $U_\mu=\exp(i\mu P Q_2)$, and we measure with the second probe. The first instance results in the effective observable $\Eo^{(\lambda,0)}$, and the second $\Fo^{(0,\mu)}$. These observables have a form similar to those in \eqref{eq:marginala} and \eqref{eq:marginalb}:

\begin{equation}
\begin{split}
	\Eo^{(\lambda,0)} (X) = (\chi_X * e^{(\lambda,0)})(Q),\\
	\Fo^{(0,\mu)} (Y) =  (\chi_Y * f^{(0,\mu)})(P).\label{eq:indiv}
\end{split}
\end{equation}

By using Eqs. \eqref{eq:marginala}, \eqref{eq:marginalb} and \eqref{eq:indiv}, the disturbance given by Di Lorenzo $\Delta_{\text{DL}}$ is the difference in the variances of the measurement statistics for the marginal observable and its individual measurement counterpart with regard to some system state $\psi$, i.e.,
\begin{subequations}
\begin{align}
	\begin{split}
		\Delta_{\text{DL}}(Q)&=\var{\Eo^{(\lambda,\mu)},\psi} - \var{\Eo^{(\lambda,0)},\psi} \\
		&= \var{e^{(\lambda,\mu)}}-\var{e^{(\lambda,0)}},
	\end{split}\\
	\begin{split}
		\Delta_{\text{DL}}(P)&=\var{\Fo^{(\lambda,\mu)},\psi} - \var{\Fo^{(0,\mu)},\psi} \\
		&= \var{f^{(\lambda,\mu)}}-\var{f^{(0,\mu)}}.
	\end{split}
\end{align}
\end{subequations}

Note that the state-dependent parts vanish due to the additive nature of the variance of convolutions. If we use a general correlated state $\sigma_{12}$ for our probes, then these errors take the form
\begin{align}
\begin{split}
	\Delta_{\text{DL}}(Q) =& \frac{(1-\kappa)^2}{4}\mu^2\,\var{Q_2,\sigma_{12}}\\
	 &- \frac{(1-\kappa)}{\lambda}\mu\,\cov{Q_1,Q_2,\sigma_{12}},\label{eq:DLQ}
\end{split}\\
\begin{split}
	\Delta_{\text{DL}}(P) =& \frac{(1+\kappa)^2}{4}\lambda^2\,\var{P_1,\sigma_{12}}\\
	 &- \frac{(1+\kappa)}{\mu}\lambda\,\cov{P_1,P_2,\sigma_{12}},\label{eq:DLP}
\end{split}
\end{align}
where $\cov{A,B,\rho}$ denotes the covariance between the two observables with regards to the state $\rho$: $\cov{A,B,\rho} = \tr{\tfrac{1}{2}(AB+BA)\rho}-\tr{A\rho}\tr{B\rho}$. 



If we begin with the states of our probes being uncorrelated, i.e., $\sigma_{12}=\sigma_1\otimes\sigma_2$, then the covariance terms in $\Delta_{\text{DL}}(Q)$ and $\Delta_{\text{DL}}(P)$ vanish, and the measures are strictly non-negative. However, the inclusion of correlated probe states in the model means that the covariance terms can be nonzero and, indeed, can in some instances be large enough that $\Delta_{\text{DL}}$ is negative. Interpreting this negativity as indicating the absence of disturbance would be implausible: {\em any} nonzero value of $\Delta_{\text{DL}}$ indicates an influence of the other measurement. The occurrence of negative values is not surprising, however, since Di Lorenzo begins by calibrating with a poor reference measurement; had he chosen the ideal reference measurement, then his disturbance value would coincide with the random error contribution in \eqref{eq:e2}, and he would have recovered \eqref{eq:ur}.

By using correlated probe states we introduce the concept of focusing, where $\Eo^{(\lambda,\mu)}$ ($\Fo^{(\lambda,\mu)}$) is a more precise approximation of position (momentum) than $\Eo^{(\lambda,0)}$ ($\Fo^{(0,\mu)}$), despite being the marginal of a joint observable. Note that all that matters for this focusing to be able to occur is the existence of some initial correlation between the probes, and in this sense entanglement is no more significant than being able to prepare the probes in a mixed state.

In the next section we show two examples in which our model can lead to both $\Delta_{\text{DL}}(Q)$ and $\Delta_{\text{DL}}(P)$ being negative simultaneously. In such cases we have a setup in which a joint measurement of position and momentum is more precise than if we separately performed individual measurements as described above. In what follows, we relabel $\Delta_{\text{DL}}$ by $\mathcal{F}$, so $\mathcal{F}(Q):=\Delta_{\text{DL}}(Q)$, etc.

\emph{Examples of focusing.}\quad The first case we will consider is that of probes prepared in a pure entangled state
\cite{Note4}
. In particular, we choose the unbiased two-mode Gaussian state, given in the position representation by
\begin{equation}
	\varphi_{12}(q_1,q_2) = \left(\frac{4\det D}{\pi^2}\right)^{1/4} \exp[-(q_1,q_2)D(q_1,q_2)^T],
\end{equation}
where $D$ is the positive-definite matrix
\begin{align}
	D = 
	\begin{pmatrix}
		a & b\\
		b & d
	\end{pmatrix} &= 
	\begin{pmatrix}
		\avg{P_1^2}_{\varphi_{12}} & \avg{P_1P_2}_{\varphi_{12}} \\
		\avg{P_1P_2}_{\varphi_{12}} & \avg{P_2^2}_{\varphi_{12}}
	\end{pmatrix}\\
	&= 4\det D
	\begin{pmatrix}
		\avg{Q_2^2}_{\varphi_{12}} & -\avg{Q_1Q_2}_{\varphi_{12}} \\
		-\avg{Q_1Q_2}_{\varphi_{12}} & \avg{Q_1^2}_{\varphi_{12}}
	\end{pmatrix}.\notag
\end{align}
Here we have used the shorthand $\avg{A}_\psi = \braket{\psi}{A\psi}$. Since $\varphi_{12}$ is unbiased, i.e., $\avg{Q_i}_{\varphi_{12}}=\avg{P_i}_{\varphi_{12}}=0$ for $i=1,2$, the variances and covariances reduce: $\var{Q_i,\varphi_{12}}=\avg{Q_i^2}_{\varphi_{12}}$ and $\cov{Q_1,Q_2,\varphi_{12}}=\avg{Q_1Q_2}_{\varphi_{12}}$, etc. With this in mind, Eqs. \eqref{eq:DLQ} and \eqref{eq:DLP} can be expressed in terms of the components of $D$, and so the conditions $\mathcal{F}(Q)<0$ and $\mathcal{F}(P)<0$ are equivalent to 
\begin{align}
	-\frac{(1-\kappa)}{\lambda}\mu\, b&>\frac{(1-\kappa)^2}{4}\mu^2 \,a>0,\label{eq:GaussQ}\\
	\frac{(1+\kappa)}{\mu}\lambda\, b&> \frac{(1+\kappa)^2}{4}\lambda^2 \,a>0.\label{eq:GaussP}
\end{align}
If we set $\abs{\kappa}<1$, then both $1+\kappa$ and $1-\kappa$ are positive, so from Eq. \eqref{eq:GaussQ} we find that $b<0$, while Eq. \eqref{eq:GaussP} shows that $b>0$. (In order to arrive at these inequalities we have used the fact that $\lambda,\mu>0$.) We conclude that, while using the two-mode Gaussian state $\varphi_{12}$, we are unable to achieve both $\mathcal{F}(Q)<0$ and $\mathcal{F}(P)<0$ if we set $\abs{\kappa}<1$.

If, however, $\abs{\kappa}>1$, then it is possible to attain both $\mathcal{F}(Q)<0$ and $\mathcal{F}(P)<0$. By considering the cases $\kappa>1$ and $\kappa<-1$ separately, it is quickly shown that we can achieve focusing on both approximate observables for the state $\varphi_{12}$ iff $\abs{\kappa}>1$ and
\begin{equation}
	\frac{\lambda\mu}{4}(1+\abs{\kappa})<\frac{\abs{b}}{a}.\label{eq:2mode}
\end{equation}

The second example that we consider is a mixed state $\sigma_{12}=\sigma= p\sigma_1+(1-p)\sigma_2$, where $0<p<1$. Both pure states $\sigma_i$ are product states composed of two single-mode Gaussian states centered on the point $(x_i,k_i)$ in phase space, i.e., $\sigma_i = P_{\varphi_i^{(1)}\otimes\varphi_i^{(2)}}$ where $\avg{Q_1}_{\sigma_i} = \avg{Q_2}_{\sigma_i}=x_i$ and $\avg{P_1}_{\sigma_i}=\avg{P_2}_{\sigma_i}=k_i$. Further to this, we assume that the pure states have a fixed variance $S$ with respect to both position operators and $R$ with respect to both momentum operators, i.e. $\var{Q_1,\sigma_i}=\var{Q_2,\sigma_i}=S$ and $\var{P_1,\sigma_i}=\var{P_2,\sigma_i}=R$ for both $i$. The covariances of $Q_1$, $Q_2$ and $P_1$, $P_2$ with respect to the state $\sigma$ are then
\begin{equation}
\begin{split}
	\cov{Q_1,Q_2,\sigma}&=(p-p^2)(x_1-x_2)^2,\\
	\cov{P_1,P_2,\sigma}&=(p-p^2)(k_1-k_2)^2.
\end{split}
\end{equation}
Both of these covariance terms are positive for any possible value of $p$, so for both $\mathcal{F}(Q)$ and $\mathcal{F}(P)$ to be negative, it is necessary that $\abs{\kappa}<1$; if $\abs{\kappa}>1$, then either $1+\kappa$ or $1-\kappa$ will become negative, and in order for focusing to occur for both observables, this requires the corresponding covariance term must be negative, with the other being positive. We may position these two pure states a large distance away in phase space and allow for the covariances to keep increasing. Now, \eqref{eq:DLQ} and \eqref{eq:DLP} become, respectively, 
\begin{align}
	\mathcal{F}(Q) =& \frac{(1-\kappa)^2}{4}\mu^2\,S\\
	 &+ \frac{(1-\kappa)}{\lambda}\mu\,\cov{Q_1,Q_2,\sigma_{12}}
	 \left[\frac{1-\kappa}4\lambda\mu-1  \right],\nonumber\\
	\mathcal{F}(P) =& \frac{(1+\kappa)^2}{4}\lambda^2\,R\\
	 &+ \frac{(1+\kappa)}{\mu}\lambda\,\cov{P_1,P_2,\sigma_{12}}
	 \left[\frac{1+\kappa}4\lambda\mu-1  \right];\nonumber
\end{align}
here we see that  joint focusing will occur if the covariance terms are made sufficiently large, and 
\begin{equation}
	\frac{\lambda\mu}{4}(1+\abs{\kappa})<1,
\end{equation}
which is similar to \eqref{eq:2mode}.

\emph{Conclusion.}\quad In this Letter we extended the Arthurs-Kelly model to allow for probes prepared in an arbitrary state. In doing so, we showed that the resulting effective observable measured on our system is covariant under phase space translations. The marginals of these observables satisfy the error-disturbance relation, contrary to the claims of Di Lorenzo. 

We showed Di Lorenzo's proposed measure of disturbance to actually be a measure of relative imprecision between two approximations of the ideal position or momentum observables. It is not a valid measure of disturbance, but does indicate the presence of focusing, where the marginals of a joint position and momentum measurement can be more precise than those performed separately. Focusing arises through the use of initial correlations between the probes, as was shown by examples with both entangled or separable probe states.

\emph{Acknowledgements.}\quad The authors thank Pieter Kok for numerous useful discussions up to and during the writing of this Letter. Further thanks are due to two anonymous referees, whose comments led to a number of clarifications in the presentation. The authors gratefully acknowledge support through the White Rose Studentship Network {\em Optimising Quantum Processes and Quantum Devices for future Digital Economy Applications}.

\bibliographystyle{unsrt}


\onecolumngrid
\appendix

\newpage

\section*{Supplemental material to ``Focusing in Arthurs-Kelly-type Joint Measurements with Correlated Probes''}

\section{Proof of Theorem 1}

We first assume that the probes are prepared in the arbitrary pure state $\varphi_{12}$, with the position coordinate denoted by $q_1$ for the first probe and $q_2$ for the second. We further assume the considered system is prepared in the state $\psi$ with position coordinate $q$. By using the form of the coupling unitary given in (4) and the identity $\exp(-i\lambda qP)\psi(x)=\psi(x-\lambda q)$, the combined state of the probes with the considered system is given by $\Psi$:
\begin{equation}
\begin{split}
	\Psi(q,q_1,q_2)&=U(\psi\otimes\varphi_{12})(q,q_1,q_2)\\
	&= e^{-i\lambda Q P_1}e^{i\mu P Q_2} e^{-i\tfrac{\lambda\mu}{2}(\kappa+1) P_1Q_2}(\psi\otimes\varphi_{12})(q,q_1,q_2)\\
	&= e^{-i\lambda q P_1}e^{i\mu q_2 P} e^{-i\tfrac{\lambda\mu}{2}(\kappa+1)q_2 P_1}(\psi\otimes\varphi_{12})(q,q_1,q_2)\\
	&=\psi(q+\mu q_2)\varphi_{12}(q_1-\lambda q - \tfrac{\lambda\mu}{2}(\kappa+1)q_2,q_2).
\end{split}
\end{equation}

On the first probe we measure position, whilst on the second we measure momentum, so before we calculate the effective observable, we perform a Fourier transform on the final argument of $\Psi$:
\begin{equation}
	\Psi(q,q_1,q_2)\mapsto\widetilde{\Psi}(q,q_1,p_2)=\frac{1}{\sqrt{2\pi}}\int_\R dq_2\, e^{-ip_2q_2}\Psi(q,q_1,q_2).\label{eq:Fourier}
\end{equation}
The measurements performed on the probes are ideal measurements, described by projection-valued measures; the position measurement on the first probe is denoted by $\Eo^{Q_1}$, and the momentum measurement on the second by $\Eo^{P_2}$. As is given in (7), we find the effective observable on the considered system $\Go^{(\lambda,\mu)}$ via
\begin{equation}
		\braket{\psi}{\Go^{(\lambda,\mu)}(X \times Y)\psi}\equiv \braket{\Psi}{I\otimes \Eo^{Q_1}(\lambda X)\otimes \Eo^{P_2}(\mu Y)\Psi}\label{eq:prob-rep2}.
\end{equation}
(The reason for the insertion of the scaling parameters $\lambda$, $\mu$ will become clear in \eqref{eq:Emarg} and \eqref{eq:Fmarg}.)
By making use of \eqref{eq:Fourier}, we calculate the form of $\Go^{(\lambda,\mu)}$:
\begin{equation}
\begin{split}
	\braket{\psi}{\Go^{(\lambda,\mu)}(X \times Y)\psi} =&\frac{1}{2\pi}\int_{\R^8}dq\,dq'\,dq_1\,dq_1'\,dq_2\,dq_2'\,dp_2\,dp_2'\, e^{i(p_2'q_2'-p_2q_2)}\overline{\psi(q'+\mu q_2')\varphi_{12}(q_1'-\lambda q' - \tfrac{\lambda\mu}{2}(\kappa+1)q_2',q_2')}\\
	&\qquad\quad\times \psi(q+\mu q_2)\varphi_{12}(q_1-\lambda q - \tfrac{\lambda\mu}{2}(\kappa+1)q_2,q_2)\braket{q'}{q}\braket{q_1'}{\Eo^{Q_1}(\lambda X)q_1}\braket{p_2'}{\Eo^{P_2}(\mu Y)p_2}.\label{eq:full_expression}
\end{split}
\end{equation}
After expressing $\Eo^{Q_1}(\lambda X)$ and $\Eo^{P_2}(\mu Y)$ in terms of pseudo-eigenvectors of $Q_1$ and $P_2$, respectively
\begin{subequations}
	\begin{align}
		\Eo^{Q_1}(\lambda X) &= \int_{\lambda X} dq_1''\, \ketbra{q_1''}{q_1''} = \lambda \int_X dq_1'' \, \ketbra{\lambda q_1''}{\lambda q_1''};\\
		\Eo^{P_2}(\mu Y) &= \int_{\mu Y} dp_2''\, \ketbra{p_2''}{p_2''} = \mu \int_Y dp_2'' \, \ketbra{\mu p_2''}{\mu p_2''},
	\end{align}
\end{subequations}
the right hand side of \eqref{eq:full_expression} reduces to
\begin{equation}
\begin{split}
	\braket{\psi}{\Go^{(\lambda,\mu)}(X \times Y)\psi} =& \frac{\lambda\mu}{2\pi}\int_{X \times Y} dq_1''\,dp_2''\, \int_{\R^4} dq\,dq'\,dq_2\,dq_2'e^{-i\mu p_2''(q_2-q_2')}\overline{\psi(q'+\mu q_2')\varphi_{12}(\lambda(q_1''-q' - \tfrac{\mu}{2}(\kappa+1)q_2'),q_2')}\\
	&\qquad\qquad\qquad\qquad\qquad\qquad\qquad\qquad\qquad\times\psi(q+\mu q_2)\varphi_{12}(\lambda(q_1''-q - \tfrac{\mu}{2}(\kappa+1)q_2),q_2)\braket{q'}{q}\\
	=& \int_{X \times Y} dq_1\,dp_2\, \left(\sqrt{\frac{\lambda\mu}{2\pi}} \int_{\R^2} dq\,dq_2\,e^{-i\mu p_2q_2}\psi(q+\mu q_2)\varphi_{12}(\lambda(q_1-q - \tfrac{\mu}{2}(\kappa+1)q_2),q_2) \ket{q}\right)^*\\
	&\qquad\qquad\qquad\times  \sqrt{\frac{\lambda\mu}{2\pi}} \int_{\R^2} dq\,dq_2\,e^{-i\mu p_2q_2}\psi(q+\mu q_2)\varphi_{12}(\lambda(q_1-q - \tfrac{\mu}{2}(\kappa+1)q_2),q_2) \ket{q}.
\end{split}
\end{equation}
We define $q'=q+\mu q_2$, so $q_2=\tfrac{1}{\mu}(q'-q)$, $dq_2'=\tfrac{1}{\mu}dq'$ and $q+\tfrac{\mu}{2}(\kappa+1)q_2=\tfrac{1}{2}\big((1-\kappa)q+(1+\kappa)q'  \big)$. Therefore
\begin{equation}
	\begin{split}
		\braket{\psi}{\Go^{(\lambda,\mu)}(X \times Y)\psi} =&\int_{X \times Y} dq_1\,dp_2\, \left(\sqrt{\frac{\lambda}{2\pi\mu}} \int_{\R^2} dq\,dq'\,e^{i p_2(q-q')}\psi(q')\varphi_{12}\big(\lambda\big(q_1-\tfrac{1}{2}\big((1-\kappa)q+(1+\kappa)q'  \big)\big),\tfrac{1}{\mu}(q'-q)\big) \ket{q}\right)^*\\
		&\qquad\qquad\qquad \times\sqrt{\frac{\lambda}{2\pi\mu}} \int_{\R^2} dq\,dq'\,e^{i p_2(q-q')}\psi(q')\varphi_{12}\big(\lambda\big(q_1-\tfrac{1}{2}\big((1-\kappa)q+(1+\kappa)q'  \big)\big),\tfrac{1}{\mu}(q'-q)\big) \ket{q}\\
		=& \int_{X \times Y} dq_1\,dp_2\, \left( \int_{\R^2} dq\,dq'\, K_{q_1p_2}(q,q')\psi(q')\ket{q} \right)^*\int_{\R^2} dq\,dq'\, K_{q_1p_2}(q,q')\psi(q')\ket{q}\\
		=&\braket{\psi}{\left(\int_{X\times Y}dq_1\,dp_2\,K_{q_1p_2}^*K_{q_1p_2}\right)\psi}.
	\end{split}
\end{equation}
We have therefore found our effective observable:
\begin{equation}
	\Go^{(\lambda,\mu)}(X\times Y)=\int_{X\times Y}dq\,dp\, K_{qp}^*K_{qp},
\end{equation}
where $K_{qp}$ has the kernel
\begin{equation}
	K_{qp}(x,x')=e^{ip(x-x')}\varphi_{12}\big(\lambda\big(q-\tfrac{1}{2}\big((1-\kappa)x+(1+\kappa)x'\big)\big),\tfrac{1}{\mu}(x'-x)\big).
\end{equation}
$K_{qp}$ can be rewritten in the following way:
\begin{equation}
\begin{split}
	K_{qp} &= \int_{\R^2}dx\,dx'\,e^{ip(x-x')}\varphi_{12}\big(\lambda\big(q-\tfrac{1}{2}\big((1-\kappa)x+(1+\kappa)x'\big)\big),\tfrac{1}{\mu}(x'-x)\big)\ketbra{x}{x'}\\
	&=  \int_{\R^2}dx\,dx'\,e^{ip((x-q)-(x'-q))}\varphi_{12}\big(\lambda\big(-\tfrac{1}{2}\big((1-\kappa)(x-q)+(1+\kappa)(x'-q)\big)\big),\tfrac{1}{\mu}((x'-q)-(x-q))\big)\ketbra{x}{x'}\\
	&=  \int_{\R^2}dx\,dx'\,e^{ip(x-x')}\varphi_{12}\big(\lambda\big(-\tfrac{1}{2}\big((1-\kappa)x+(1+\kappa)x'\big)\big),\tfrac{1}{\mu}(x'-x)\big)\ketbra{x+q}{x'+q}\\
	&=  e^{-iqP}\left(\int_{\R^2}dx\,dx'\,e^{ip(x-x')}\varphi_{12}\big(\lambda\big(-\tfrac{1}{2}\big((1-\kappa)x+(1+\kappa)x'\big)\big),\tfrac{1}{\mu}(x'-x)\big)\ketbra{x}{x'}\right)e^{iqP}\\
	&=  e^{-iqP}e^{ipQ}\left(\int_{\R^2}dx\,dx'\,\varphi_{12}\big(\lambda\big(-\tfrac{1}{2}\big((1-\kappa)x+(1+\kappa)x'\big)\big),\tfrac{1}{\mu}(x'-x)\big)\ketbra{x}{x'}\right)e^{-ipQ}e^{iqP}\\
	&=  W_{qp}K_{00}W_{qp}^*,
\end{split}
\end{equation}
where $W_{qp}=\exp[-i(qP-pQ)]$ is the Weyl operator, which generates translations in phase space. From this we see
\begin{equation}
\begin{split}
	W_{qp}\Go^{(\lambda,\mu)}(Z)W_{qp}^* &= \int_Z dq'\,dp'\, W_{qp}K_{q'p'}^*K_{q'p'}W_{qp}^*\\
	&= \int_Z dq'\,dp'\, (W_{qp}K_{q'p'}W_{qp}^*)^* (W_{qp}K_{q'p'}W_{qp}^*)\\
	&= \int_{Z+(q,p)} dq'\,dp'\, K_{q'p'}^*K_{q'p'}\\
	&=\Go^{(\lambda,\mu)}(Z+(q,p)).
\end{split}
\end{equation}
We have thus shown that if we start with probes prepared in an arbitrary pure state, the effective observable measured on the system is covariant under phase space translations.

Next, we consider the case of mixed states $\sigma=\sum_i p_i \sigma_i$, where the $\sigma_i$ are arbitrary pure states. The post-coupling state is now given by $U(P_\psi\otimes\sigma)U^* = \sum_i p_i U(P_\psi\otimes\sigma_i)U^*$, and the effective observable is now found as follows:
\begin{equation}
\begin{split}
	\braket{\psi}{\Go^{(\lambda,\mu)}(X\times Y)\psi} &= \tr{U(P_\psi\otimes\sigma)U^* (I\otimes \Eo^{Q_1}(\lambda X) \otimes \Eo^{P_2}(\mu Y))}\\
	&= \sum_i p_i\,\tr{U(P_\psi\otimes\sigma_i)U^* (I\otimes \Eo^{Q_1}(\lambda X) \otimes \Eo^{P_2}(\mu Y))}\\
	&= \braket{\psi}{ \left(\sum_i p_i \Ho_i^{(\lambda,\mu)}(X\times Y)\right)\psi},
\end{split}
\end{equation}
where $\Ho_i^{(\lambda,\mu)}$ is the covariant phase space observable associated with the probes prepared in the pure state $\sigma_i$. From this we quickly find that
\begin{equation}
\begin{split}
	W_{qp}\Go^{(\lambda,\mu)}(Z)W_{qp}^*&= \sum_i p_i\, W_{qp}\Ho_i^{(\lambda,\mu)}(Z)W_{qp}^*\\
	&= \sum_i p_i \Ho_i^{(\lambda,\mu)}(Z+(q,p))\\
	&= \Go^{(\lambda,\mu)}(Z+(q,p)),
\end{split}
\end{equation}
proving the covariance, and hence Theorem 1.\\

\section{The Marginals of \texorpdfstring{$\Go^{(\lambda,\mu)}$}{G}}

We find the marginals of the observable $\Go^{(\lambda,\mu)}$, $\Eo^{(\lambda,\mu)}$ and $\Fo^{(\lambda,\mu)}$, by integrating over the outcome space of the other variable (this may be seen as projecting down to a one-dimensional subspace of phase space):
\begin{subequations}
	\begin{align}
		\Eo^{(\lambda,\mu)}(X)&=\Go^{(\lambda,\mu)}(X\times\R);\\
		\Fo^{(\lambda,\mu)}(Y)&=\Go^{(\lambda,\mu)}(\R\times Y).
	\end{align}
\end{subequations}
Considering the case where the probes are prepared in the pure state $\varphi_{12}$, we first calculate $\Eo^{(\lambda,\mu)}$:
\begin{equation}
\begin{split}
	\Eo^{(\lambda,\mu)}(X) =&\int_{X\times\R}dq\,dp\,K_{qp}^*K_{qp}\\
	 =&\frac{\lambda}{\mu} \int_{X} dq\, \int_{\R^3} dx\, dx'\, dy'\,\left(\frac{1}{2\pi}\int_\R dp\, e^{ip(y'-x')}\right)\overline{\varphi_{12}(\lambda(q-\tfrac{1}{2}((1-\kappa)x+(1+\kappa)y')),\tfrac{1}{\mu}(y'-x))}\\
			&\qquad\qquad\qquad\qquad\qquad\qquad\qquad\times\varphi_{12}(\lambda(q-\tfrac{1}{2}((1-\kappa)x+(1+\kappa)x')),\tfrac{1}{\mu}(x'-x))\ketbra{y'}{x'}.\\
			=&\frac{\lambda}{\mu} \int_{X} dq\, \int_{\R^2} dx\, dx'\,\abs{\varphi_{12}(\lambda(q-\tfrac{1}{2}((1-\kappa)x+(1+\kappa)x')),\tfrac{1}{\mu}(x'-x))}^2 \ketbra{x'}{x'},
\end{split}
\end{equation}
where we have used the identity $\int_\R dk\, \exp(ikx)=2\pi \delta(x)$. We define $q'=x'-x$, so $x=x'-q'$, $dx=-dq'$ and $(1-\kappa)x+(1+\kappa)x'=2x'-(1-\kappa)q'$. $\Eo^{(\lambda,\mu)}$ then takes the form:
\begin{equation}
\begin{split}
	\Eo^{(\lambda,\mu)}(X)&= \frac{\lambda}{\mu} \int_{X} dq\, \int_{\R^2} dq'\, dx'\,\abs{\varphi_{12}(\lambda(\tfrac{1}{2}(1-\kappa)q'-(x'-q)),\tfrac{1}{\mu}q')}^2 \ketbra{x'}{x'}\\
	&=\int_X dq\,\int_\R dx' e^{(\lambda,\mu)}(x'-q)\ketbra{x'}{x'}\\
	&=\int_\R dq\, \chi_X(q)e^{(\lambda,\mu)}(Q-q)\\
	&=(\chi_X * e^{(\lambda,\mu)})(Q),\label{eq:Emarg}
\end{split}
\end{equation}
as is given in (9a). The probability distribution $e^{(\lambda,\mu)}$, which characterizes the noise in the measurement of $\Eo^{(\lambda,\mu)}$, is of the form
\begin{equation}
	e^{(\lambda,\mu)}(q) =\frac{\lambda}{\mu}\int_\R dq'\, \abs{\varphi_{12}(\lambda(\tfrac{1}{2}(1-\kappa)q'-q),\tfrac{1}{\mu}q')}^2,
\end{equation}
with first and second moments
\begin{align}
\begin{split}
	e^{(\lambda,\mu)}[1]&=\int_\R dq\, q\, e^{(\lambda,\mu)}(q) \\
	&= \frac{1}{\mu}\int_{\R^2}dq\,dq'(\tfrac{1}{2}(1-\kappa)q'-\tfrac{1}{\lambda}q)\abs{\varphi_{12}\big(q,\tfrac{1}{\mu}q'\big)}^2\\
	&=\int_{\R^2}dq\,dq'(\tfrac{\mu}{2}(1-\kappa)q'-\tfrac{1}{\lambda}q)\abs{\varphi_{12}(q,q')}^2\\
	&=\frac{\mu}{2}(1-\kappa)\avg{Q_2}_{\varphi_{12}}-\frac{1}{\lambda}\avg{Q_1}_{\varphi_{12}};
\end{split}\label{eq:efirst}\\
\begin{split}
	e^{(\lambda,\mu)}[2]&=\int_\R dq\, q^2\, e^{(\lambda,\mu)}(q)\\
	&=  \int_{\R^2}dq\,dq'(\tfrac{\mu}{2}(1-\kappa)q'-\tfrac{1}{\lambda}q)^2\abs{\varphi_{12}(q,q')}^2\\
&= \frac{\mu^2}{4}(1-\kappa)^2 \avg{Q_2^2}_{\varphi_{12}}+\frac{1}{\lambda^2}\avg{Q_1^2}_{\varphi_{12}}-\frac{\mu}{\lambda}\avg{Q_1Q_2}_{\varphi_{12}},\label{eq:esecond}
\end{split}
\end{align}
where $\avg{Q_1}_{\varphi_{12}}=\braket{\varphi_{12}}{Q_1\varphi_{12}}$, etc. Using \eqref{eq:efirst} and \eqref{eq:esecond}, the variance of $e^{(\lambda,\mu)}$ is
\begin{equation}
	\var{e^{(\lambda,\mu)}}= \frac{1}{\lambda^2}\var{Q_1,\varphi_{12}}+\frac{\mu^2}{4}(1-\kappa)^2\var{Q_2,\varphi_{12}}-\frac{\mu}{\lambda}(1-\kappa)\cov{Q_1,Q_2,\varphi_{12}},\label{eq:evar}
\end{equation}
where $\cov{Q_1,Q_2,\varphi_{12}}=\avg{Q_1Q_2}_{\varphi_{12}}-\avg{Q_1}_{\varphi_{12}}\avg{Q_2}_{\varphi_{12}}$ is the covariance of $Q_1$ and $Q_2$ with respect to $\varphi_{12}$. In a similar fashion, we derive an explicit form for $\Fo^{(\lambda,\mu)}$, the first step of which is to perform a Fourier transform on $\varphi_{12}$:
\begin{subequations}
\begin{align}
		\varphi_{12}(\lambda(q-\tfrac{1}{2}((1-\kappa)x+(1+\kappa)x')),\tfrac{1}{\mu}(x'-x))=&\frac{1}{2\pi}\int_{\R^2}dw\,dz\, e^{iw(q-\tfrac{1}{2}((1-\kappa)x+(1+\kappa)x'))}e^{iz(x'-x)}\widetilde{\varphi}_{12}(\lambda w,\tfrac{z}{\mu}),\\
		\overline{\varphi_{12}(\lambda(q-\tfrac{1}{2}((1-\kappa)x+(1+\kappa)y')),\tfrac{1}{\mu}(y'-x))}=&\frac{1}{2\pi}\int_{\R^2}dw'\,dz'\, e^{-iw'(q-\tfrac{1}{2}((1-\kappa)x+(1+\kappa)y'))}e^{-iz'(y'-x)}\overline{\widetilde{\varphi}_{12}(\lambda w',\tfrac{z'}{\mu})},
\end{align}
\end{subequations}
and so we find
\begin{equation}
\begin{split}
	\mathsf{F}^{(\lambda,\mu)}(Y) =& \frac{\lambda}{2\pi\mu} \int_{\R \times Y} dq\,dp\, \int_{\R^3} dx\, dx'\, dy'\, e^{ip(y'-x')} \int_{\R^4}\frac{dw\,dw'\,dz\,dz'}{4\pi^2}e^{iq(w-w')}e^{-\tfrac{iw}{2}((1-\kappa)x+(1+\kappa)x')}\\
		&\qquad\qquad\qquad\qquad\times e^{\tfrac{iw'}{2}((1-\kappa)x+(1+\kappa)y')}e^{ix(z'-z)}e^{izx'}e^{-iz'y'}\widetilde{\varphi}_{12}(\lambda w,\tfrac{z}{\mu})\overline{\widetilde{\varphi}_{12}(\lambda w',\tfrac{z'}{\mu})}\ketbra{y'}{x'}\\
		=& \frac{\lambda}{2\pi\mu} \int_{Y} dp\, \int_{\R^3} dx\, dx'\, dy'\, e^{ip(y'-x')} \int_{\R^4}\frac{dw\,dw'\,dz\,dz'}{2\pi}\left(\frac{1}{2\pi}\int_\R dq\,e^{iq(w-w')}\right)e^{-\tfrac{iw}{2}((1-\kappa)x+(1+\kappa)x')}\\
		&\quad\qquad\qquad\qquad\qquad\qquad\times e^{\tfrac{iw'}{2}((1-\kappa)x+(1+\kappa)y')}e^{ix(z'-z)}e^{izx'}e^{-iz'y'}\widetilde{\varphi}_{12}(\lambda w,\tfrac{z}{\mu})\overline{\widetilde{\varphi}_{12}(\lambda w',\tfrac{z'}{\mu})}\ketbra{y'}{x'}\\
		=& \frac{\lambda}{2\pi\mu} \int_{Y} dp\,  \int_{\R^3}dw\,dz\,dz'\, \int_{\R^2} dx'\, dy'\, e^{i(p+\tfrac{w}{2}(1+\kappa))(y'-x')}\left(\frac{1}{2\pi}\int_\R dx\,e^{ix(z'-z)}\right)\\
		&\qquad\qquad\qquad\qquad\qquad\qquad\qquad\qquad\times e^{izx'}e^{-iz'y'}\widetilde{\varphi}_{12}(\lambda w,\tfrac{z}{\mu})\overline{\widetilde{\varphi}_{12}(\lambda w,\tfrac{z'}{\mu})}\ketbra{y'}{x'}\\
		=& \frac{\lambda}{2\pi\mu} \int_{Y} dp\, \int_{\R^2} dx'\, dy'\,  \int_{\R^2}dw\,dz\,e^{i(p+\tfrac{w}{2}(1+\kappa)-z)(y'-x')}\abs{\widetilde{\varphi}_{12}(\lambda w,\tfrac{z}{\mu})}^2\ketbra{y'}{x'}\\
		=&\frac{\lambda}{\mu}\int_Y dp\int_{\R^2}dw\,dz\,\abs{\widetilde{\varphi}_{12}(\lambda w,\tfrac{z}{\mu})}^2 \left(\frac{1}{\sqrt{2\pi}} \int_\R dy'\, e^{iy'(p+\tfrac{w}{2}(\kappa+1)-z)}\ket{y'} \right) \left(\frac{1}{\sqrt{2\pi}} \int_\R dx'\, e^{ix'(p+\tfrac{w}{2}(\kappa+1)-z)}\ket{x'} \right)^*\\
		=&\frac{\lambda}{\mu}\int_Y dp\int_{\R^2}dw\,dz\,\abs{\widetilde{\varphi}_{12}(\lambda w,\tfrac{z}{\mu})}^2 \ketbra{p+\tfrac{w}{2}(\kappa+1)-z}{p+\tfrac{w}{2}(\kappa+1)-z}.	
\end{split}
\end{equation}
Defining $p'=p+\tfrac{w}{2}(\kappa+1)-z$, so $z=p-p'+\tfrac{w}{2}(\kappa+1)$ and $dz=-dp'$, $\Fo^{(\lambda,\mu)}$ takes the form
\begin{equation}
\begin{split}
	\Fo^{(\lambda,\mu)}(Y)&=\frac{\lambda}{\mu}\int_Y dp\int_{\R^2}dw\,dp'\,\abs{\widetilde{\varphi}_{12}\big(\lambda w,\tfrac{1}{\mu}(p-p'+\tfrac{w}{2}(\kappa+1))\big)}^2 \ketbra{p'}{p'}\\
	&=\int_Y dp\int_{\R}dp'\,f^{(\lambda,\mu)}(p'-p) \ketbra{p'}{p'}\\
	&=\int_\R dp\,\chi_Y(p)f^{(\lambda,\mu)}(P-p)\\
	&=(\chi_Y * f^{(\lambda,\mu)})(P),\label{eq:Fmarg}
\end{split}
\end{equation}
as is given in (9b). The probability distribution $f^{(\lambda,\mu)}$ is of the form
\begin{equation}
	f^{(\lambda,\mu)}(p) = \frac{\lambda}{\mu}\int_\R dw\,\abs{\widetilde{\varphi}_{12}\big(\lambda w,\tfrac{1}{\mu}(\tfrac{w}{2}(\kappa+1)-p)\big)}^2.
\end{equation}
By using the identity
\begin{equation}
	\widetilde{\varphi}_{12}\big(\lambda p, \tfrac{1}{\mu}p'\big)= \frac{\mu}{\lambda}\widetilde{\varphi}_{12}(\tfrac{1}{\lambda}p,\mu p'),
\end{equation}
and following the same method used to derive \eqref{eq:efirst} and \eqref{eq:esecond}, we find the first and second moments of $f^{(\lambda,\mu)}$:
\begin{align}
	f^{(\lambda,\mu)}[1]&= \frac{\lambda}{2}(1+\kappa)\avg{P_1}_{\varphi_{12}}-\frac{1}{\mu}\avg{P_2}_{\varphi_{12}};\label{eq:ffirst}\\
	f^{(\lambda,\mu)}[2]&=\frac{\lambda^2}{4}(1+\kappa)^2\avg{P_1^2}_{\varphi_{12}}+\frac{1}{\mu^2}\avg{P_2^2}_{\varphi_{12}}-\frac{\lambda}{\mu}(1+\kappa)\avg{P_1P_2}_{\varphi_{12}}.\label{eq:fsecond}
\end{align}
From these, the variance of $f^{(\lambda,\mu)}$ is given by
\begin{equation}
	\var{f^{(\lambda,\mu)}}=\frac{\lambda^2}{4}(1+\kappa)^2\var{P_1,\varphi_{12}}+\frac{1}{\mu^2}\var{P_2,\varphi_{12}}-\frac{\lambda}{\mu}(1+\kappa)\cov{P_1,P_2,\varphi_{12}}.\label{eq:fvar}
\end{equation}

\begin{remark}
	It is now clear why we use the scaled sets $\lambda X$ and $\mu Y$ in (7) and \eqref{eq:prob-rep2}: the scaling is such that the marginal observables $\Eo^{(\lambda,\mu)}$ and $\Fo^{(\lambda,\mu)}$ are direct smearings of position and momentum, rather than of scaled versions.
\end{remark}
We will now return to the case of the mixed state $\sigma=\sum_i p_i \sigma_i$, where $\sigma_i=P_{\varphi_i}$ are arbitrary pure states. The marginals of the effective observable $\Go^{(\lambda,\mu)}$ are now given in terms of the marginals of the effective observables derived from $\sigma_i$:
\begin{subequations}
\begin{align}
\begin{split}
	\Eo^{(\lambda,\mu)} (X) &= \Go^{(\lambda,\mu)}(X\times \R) =\sum_i p_i \Ho^{(\lambda,\mu)}_i(X\times \R) = \sum_i p_i \Mo_i^{(\lambda,\mu)}(X)\\
	&=\sum_i p_i (\chi_X * m_i^{(\lambda,\mu)})(Q)=(\chi_X*e^{(\lambda,\mu)})(Q),
\end{split}\\
\begin{split}
	\Fo^{(\lambda,\mu)} (Y) &= \Go^{(\lambda,\mu)}(\R\times Y) =\sum_i p_i \Ho^{(\lambda,\mu)}_i(\R\times Y) = \sum_i p_i \No_i^{(\lambda,\mu)}(Y)\\
	&=\sum_i p_i (\chi_Y * n_i^{(\lambda,\mu)})(P)=(\chi_Y*f^{(\lambda,\mu)})(P).
\end{split}
\end{align}
\end{subequations}
As such, these marginals again have the form (9a), (9b), with the probability distributions 
\begin{subequations}
\begin{align}
	e^{(\lambda,\mu)}(q)&=\sum_i p_i m^{(\lambda,\mu)}_i(q) =\frac{\lambda}{\mu}\sum_i p_i \int_\R dq'\, \abs{\varphi_i(\lambda(\tfrac{1}{2}(1-\kappa)q'-q),\tfrac{1}{\mu}q')}^2,\\
	f^{(\lambda,\mu)}(p)&=\sum_i p_i n^{(\lambda,\mu)}_i(p) =\frac{\lambda}{\mu}\sum_i p_i \int_\R dw\,\abs{\widetilde{\varphi}_i\big(\lambda w,\tfrac{1}{\mu}(\tfrac{w}{2}(\kappa+1)-p)\big)}^2.
\end{align}
\end{subequations}
From here, and by using equations \eqref{eq:efirst}, \eqref{eq:esecond}, \eqref{eq:ffirst}, \eqref{eq:fsecond}, the first and second moments of these distributions can be readily calculated:
\begin{align}
\begin{split}
	e^{(\lambda,\mu)}[1]&=\int_\R dq\, q\, e^{(\lambda,\mu)} = \sum_i p_i \int_\R dq\, q\, m_i^{(\lambda,\mu)}(q) = \sum_i p_i m^{(\lambda,\mu)}[1] \\
	&= \sum_i p_i \left(\frac{\mu}{2}(1-\kappa)\avg{Q_2}_{\varphi_i}-\frac{1}{\lambda}\avg{Q_1}_{\varphi_i}\right) \\
	&= \frac{\mu}{2}(1-\kappa)\avg{Q_2}_{\sigma}-\frac{1}{\lambda}\avg{Q_1}_{\sigma},
\end{split}\\
	e^{(\lambda,\mu)}[2] &= \frac{\mu^2}{4}(1-\kappa)^2 \avg{Q_2^2}_{\sigma}+\frac{1}{\lambda^2}\avg{Q_1^2}_{\sigma}-\frac{\mu}{\lambda}\avg{Q_1Q_2}_{\sigma},\\
	f^{(\lambda,\mu)}[1]&= \frac{\lambda}{2}(1+\kappa)\avg{P_1}_{\sigma}-\frac{1}{\mu}\avg{P_2}_{\sigma},\\
	f^{(\lambda,\mu)}[2]&=\frac{\lambda^2}{4}(1+\kappa)^2\avg{P_1^2}_{\sigma}+\frac{1}{\mu^2}\avg{P_2^2}_{\sigma}-\frac{\lambda}{\mu}(1+\kappa)\avg{P_1P_2}_{\sigma},
\end{align}
and so these probability distributions have variances of the form given in \eqref{eq:evar} and \eqref{eq:fvar}
\begin{align}
	\var{e^{(\lambda,\mu)}}&= \frac{1}{\lambda^2}\var{Q_1,\sigma}+\frac{\mu^2}{4}(1-\kappa)^2\var{Q_2,\sigma}-\frac{\mu}{\lambda}(1-\kappa)\cov{Q_1,Q_2,\sigma},\\
	\var{f^{(\lambda,\mu)}}&=\frac{\lambda^2}{4}(1+\kappa)^2\var{P_1,\sigma}+\frac{1}{\mu^2}\var{P_2,\sigma}-\frac{\lambda}{\mu}(1+\kappa)\cov{P_1,P_2,\sigma}.
\end{align}
Note that even if we had specified that the pure states $\sigma_i$ were product states, we would still find the covariance terms appearing as a result of the classical correlations between them.\\

\section{Error values for $\Eo^{(\lambda,\mu)}$ and $\Fo^{(\lambda,\mu)}$}

In this section, we briefly review the definitions of error presented by Ozawa \cite{Ozawa2004} and Busch, Lahti and Werner \cite{BLW2013c}. While these definitions take different approaches, the purpose of this section is to show explicitly that in the case of observables of the form $\Qo_m$, where $\Qo_m(X)=(\chi_X*m)(Q)$ with $m$ a probability distribution, these error measures actually coincide. This result applies, in particular, to $\Eo^{(\lambda,\mu)}$ and $\Fo^{(\lambda,\mu)}$. (It seems useful to note this observation for future reference; the result is implicit from calculations of these quantities found in various places in the literature.)


We shall begin by discussing the definition of error given by Ozawa \cite{Ozawa2004}. In its most general terms, Ozawa's definition is an attempt to generalise root-mean-square deviations for quantum observables. Consider a system in a state $\psi$, upon which one wishes to measure the observable $\Eo^A$ with first moment operator $A=\int_\R x\, d\Eo^A(x)$. If we couple this to an auxiliary system, described by the Hilbert space $\mathcal{K}$ in a state $\xi$, with a coupling unitary $U$, and perform a sharp pointer measurement $\mathsf{Z}$ with first moment $Z$, then the error is given by
\begin{equation}
	\epsilon(\Eo,\Eo^A,\psi)^2 = \matrixel{\psi,\xi}{(U^*(I\otimes Z)U - A\otimes I)^2}{\psi,\xi},\label{eq:oz}
\end{equation}
where $\Eo$ is the effective observable given by
\begin{equation}
	\Eo(X)=\tr[\mathcal{K}]{U^*(I\otimes \mathsf{Z}(X))U (I\otimes P_\xi)}.
\end{equation}
This can be expressed in terms of quantities pertaining to the measured system alone:
\begin{equation}
\begin{split}
	\epsilon(\Eo,\Eo^A,\psi)^2=\braket{\psi}{(\Eo[1]-A)^2\psi} + \braket{\psi}{(\Eo[2]-\Eo[1]^2)\psi}.\label{eq:ozuseful}
\end{split}
\end{equation}

With \eqref{eq:ozuseful} at hand, we now return to the case we wish to consider, namely the error of $\Qo_m$ with respect to the ideal measurement $\Eo^Q$, $\epsilon(\Qo_m,\Eo^Q,\psi)$. It is easily shown that the first and second moment operators of the observable
$\Qo_\mu$ are given by
\begin{equation}
\Qo_m[1]=Q+m[1],\quad \Qo_m[2]=\Qo_m[1]^2+\var m.
\end{equation}
This gives
\begin{equation}
\epsilon(\Qo_m,\Eo^Q,\psi)^2=m[2]=m[1]^2+\var m.
\end{equation}

We next recall the definition of error introduced in \cite{BLW2013c} as an operationally meaningful quantum version of root-mean-square error. For any two probability measures $\alpha,\beta$ on $\R$ a {\it coupling} is defined to be a probability measure $\gamma$ on $\R\times \R$  with  $\alpha$ and $\beta$ as the Cartesian marginals.
The set of couplings between $\alpha$ and $\beta$ will be denoted $\Gamma(\alpha,\beta)$. Then, the (Wasserstein) 2-distance \cite{Villani} of $\alpha$ and $\beta$ is defined as
\begin{equation}\label{Wasserstein}
  \cD_2(\alpha,\beta)=  \inf_{\gamma\in\Gamma(\alpha,\beta)}  \cD^\gamma_2(\alpha,\beta)=
  \inf_{\gamma\in\Gamma(\alpha,\beta)}\left(\int |x-y|^2\,d\gamma(x,y) \right)^{\frac12}
\end{equation}
The existence of an optimal coupling is known, see \cite[Theorem 4.1]{Villani}, but it does not imply that  $\cD_2(\alpha,\beta)$  is finite. This is a distance between probability measures due to the choice of the minimizing joint probability.

We can now define the \emph{(Wasserstein) $2$-distance between observables} $\sfe,\sff$ on $\R$ using the notation $p_\rho^\sfe$, $p_\rho^\sff$ for their probability measures with respect to the state $\rho$:
\begin{equation*}
\Delta_2(\sfe,\sff):=\sup_{\rho}\cD_2(p_\rho^\sfe,p_\rho^\sff).
\end{equation*}
As a direct application of \cite[Lemma 7]{BLW2014}, one obtains
\begin{equation}
\Delta_2(\Qo_m,\Eo^Q)^2=\cD_2(m,\delta_0)^2=m[2]\equiv \epsilon(\Qo_m,\Eo^Q,\psi)^2.
\end{equation}

It is a remarkable that $\epsilon(\Qo_m,\Eo^Q,\psi)$  coincides with $\Delta_2(\Qo_m,\Eo^Q)$ considering that the former is expressly defined as a state-specific quantity while the latter represents a worst-case error measure across all states. This coincidence underpins the intuitive idea that the smearing measure $m$ characterizes the random and systematic errors in a measurement of the observable $\Qo_m$ as an approximation of $\Eo^Q$.

\bibliographystyle{unsrt}


\end{document}